# Grassroots Innovation Actors: Their Role and Positioning in Economic Ecosystems - A Comparative Study Through Complex Network Analysis.

**Marcelo S. Tedesco**[δ]     **Francisco Javier Ramos Soria**[ϕ]

**Abstract**

This study offers an examination of grassroots innovation actors and their integration within larger economic ecosystems. Through a comparative analysis in Oaxaca, Mexico; La Plata, Argentina; and Araucanía, Chile, this research sheds light on the vital role that grassroots innovation plays in broader economic ecosystems. Using Complex Network Analysis and the TE-SER model, the study unveils how these actors interact, collaborate, and influence major economic ecosystems in the context of complex social challenges. The findings highlight that actors from the grassroots innovation ecosystem make up a significant portion of the larger innovation-driven entrepreneurial economic ecosystem, accounting for between 20% and 30% in all three cases and are strategically positioned within the ecosystem's structural network. Additionally, this study emphasizes the potential for greater integration of grassroots innovation actors to leverage resources and foster socio-economic development. The research concludes by advocating for further studies in similar socio-economic contexts to enhance our understanding of integration dynamics and mutual benefits between grassroots innovation ecosystems and other larger economic systems.

**Acknowledgments:** We would like to extend our sincere gratitude for the generous support received from various institutions and individuals that made this work possible. Special thanks to the Oaxacan Institute of Entrepreneurship (IODEMC), and to Rodrigo Arnaud and Carlos Sandoval Habib for their invaluable support in data collection in Oaxaca. The work in the Araucanía would not have been possible without the funding from the Universidad de La Frontera, and we appreciate La Mesa de la Araucanía, a key actor in the region. Our gratitude also goes to Cristian Campomanes and Constanza Casanova for their comprehensive support in bringing everything together and making the work in their economic ecosystem feasible. Our appreciation extends to Gonzalo Marquez and Juan Pedro Brandi from the Universidad Nacional de La Plata for their leadership and support. Thanks to Adriana Tortajada and Eduardo Lebrija from Santander Universities, whose funding not only facilitated the international expansion of the Global Ecosystem Dynamics Initiative (GED) but also contributed to the progress of science through the validation of the models used in this work. Lastly, we appreciate Angela Sarcina from the Joint Research Centre (JRC), who encouraged us to focus on how our models and methodological approach could inform the crucial work being carried out in the economic ecosystems of Grassroot Innovation. Of course, heartfelt thanks to our entire GED team for their constant effort and tireless work in advancing knowledge.

---

[δ] Corresponding author: Marcelo S. Tedesco is the Director of Research on Economics, Society, and Management Science and the Director of the PhD program in Systemic Innovation at the Buenos Aires Institute of Technology. He is the founder and Executive Director of Global Ecosystem Dynamics (GED), an international research initiative focused on advancing the study of economic ecosystems. He also serves as an International Researcher Collaborator with MIT LIFT Lab at the Massachusetts Institute of Technology. Email: tedesco@itba.edu.ar

[ϕ] Francisco Ramos is a Research Affiliate with the Local Innovation Group at the MIT Sociotechnical Systems Research Center of the Massachusetts Institute of Technology. He is currently the Head of Data Science at Global Ecosystem Dynamics and a member of the World Economic Forum's Global Shapers Community.

# 1. Introduction

Grassroots innovation (GRI) is defined as a form of social innovation that nurtures creativity and sustainable innovation to address issues related to marginalized groups. It is characterized by engaging in activities that make use of various resources or context-appropriate technologies to foster innovations more creatively, often benefiting the community and the environment. These innovations typically emerge in response to needs, challenges, and difficulties, originating from individual users or communities in their quest to find solutions to personal or societal problems. Grassroots innovation projects have the potential to generate fresh, bottom-up solutions that tackle local problems, align with local interests and values, and resolve social, economic, and environmental challenges in marginalized communities (Prasetyo, 2016; Amir Syah et al., 2021; Feola & Nunes, 2014; Hilmi, 2012; Monaghan, 2009).

These innovations contribute to fostering democratic innovation, advocating for social justice and environmental resilience, and promoting social diversity. They represent another arena where innovation takes place, capturing non-profit economic benefits that contribute to a shift towards more sustainable consumption systems (Smith & Stirling, 2017; Dana et al., 2019; Aristizábal et al., 2016).

Grassroots innovation, frequently driven by social entrepreneurs, non-profit organizations, and the public sector, plays a pivotal transformative role. Instances such as shared mobility, community energy initiatives, and green finance exemplify its profound impact. Rooted within communities and tailored to context-specific solutions, these innovations not only address social and environmental needs but also yield significant economic benefits. Responsible innovation underscores the involvement of individuals in the innovation process, harnessing their creativity and fostering social acceptance of novelty and transformation (Fitjar et al., 2019). Nevertheless, the economic ecosystems they establish often confront distinctive challenges, as they may not always be seamlessly integrated into larger economic systems. This situation gives rise to issues related to scalability, diffusion, recognition, legitimacy, as well as access to resources and networks.

The absence of integration with more extensive economic ecosystems and the lack of supportive structures constrains the potential reach and socio-economic impact of these innovations (Radziwon, et al., 2017). Grassroots innovators frequently encounter difficulties aligning their objectives with those of broader economic ecosystems, which can result in a lack of synergy and missed opportunities (Nambisan & Baron, 2013), thereby impeding their capacity for growth and the widespread dissemination of innovative solutions.

Recognition and legitimacy represent additional crucial concerns. Grassroots innovations do not consistently receive recognition as valuable contributions to broader economic and social development. This lack of acknowledgement can hinder their ability to attract investment and support (Calvo et al., 2020), and it can prove particularly challenging when these innovations compete with more established solutions.

Access to resources and networks represents another significant challenge. Grassroots economic ecosystems often operate with limited resources and lack the necessary connections to access funding and technical knowledge (Zahra & Nambisan, 2011). This resource constraint restricts their ability to experiment and scale innovations, while the absence of robust networks hampers collaboration and mutual learning—essential elements for growth and sustainability. In addition to these challenges, grassroots innovation ecosystems encounter inherent issues related to their specific nature and context. The necessity to address local specificities and be contextually appropriate are critical aspects that must be carefully managed to ensure the relevance and effectiveness of innovations (Radziwon et al., 2017).

This paper's focus lies in describing how actors from grassroots innovation economic ecosystems integrate into larger innovation-driven economic ecosystems. Employing Complex Network Analysis, based on the understanding that every economic ecosystem is a complex system, this analysis will offer a deeper comprehension of how grassroots innovation actors assimilate into broader economic ecosystems through their collaborative relationships. It will also delineate the roles they assume within these economic ecosystems and elucidate the value they contribute to them.

## 2. Theoretical and methodological approach

### 2.1 Economic Ecosystem as Complex Systems.

To gain a deeper understanding of the social dynamics within economic ecosystems and the resulting structures, Tedesco (2022) and Tedesco et al. (2022) have proposed a theoretical approach and metrics for this purpose. According to Tedesco (2022), economic ecosystems are defined as communities of actors and individuals that interact with each other and their environment within a delimited region, shaped by its social and natural dynamics, where resources are exchanged with the function and/or purpose of creating some form of economic value. On the other hand, the term 'actor' encompasses all organizations or initiatives of an organization that exist for the benefit of the economic ecosystem to which they belong (Tedesco et al., 2018).

Examining an economic ecosystem through the lens of Complex Systems Theory can provide us with a more comprehensive understanding of its reality (Foster, 2004; Crawford et al., 2005; Meadows, 2008; Farmer, 2012; Earls, 2013; Thurner et al., 2018). This proposed approach can facilitate a rapid comprehension of how actors integrate within the economic ecosystem. Furthermore, the use of Complex Network Analysis enables a thorough investigation into the dynamics of network structure construction (Tedesco et al., 2022).

Moreover, the application of the TE-SER model (Tedesco & Serrano, 2019) allows for the description of the diversity of actors within the economic ecosystem, particularly in the case of a comprehensive examination of roles and the value contributed by actors in the grassroots innovation economic ecosystem.

Additionally, this present work considers the framework for the interrelationships among economic ecosystems proposed by Tedesco & Serrano (2019). This framework describes how, despite each economic 'sub-ecosystem' having its distinct function, as articulated by Meadows (2008) for complex systems, each of these sub-ecosystems interrelates with one another.

### 2.2 Social context of the selected economic ecosystems.

As described by Tedesco (2022), every economic ecosystem is encompassed within a social ecosystem, forming an integral and indivisible part of it. Understanding economic dynamics accurately seems impossible without considering their social dynamics, as every economic system constitutes a sub-system of the social system, which, in turn, is a sub-system of the biological system (Boulding, 1970; Bunge, 1989, Bunge, 2012, and Mobus, 2022). This perspective finds support in ecological economics, which conceptualizes economies as open subsystems within the closed biosphere, subject to biophysical laws and constraints (Daly, 1993; Victor, 2010).

Research also indicates that the economic subsystem is intricately connected with other subsystems, including the political and social domains, and the performance levels of these subsystems characterize the entire social system (Kolbin, 1985). In the bioeconomy, for example, the value of biological resources isn't solely determined by their physical characteristics; rather, it is influenced by their economic and social utilization and perception. For instance, the worth of products derived from nature can fluctuate considerably due to market demand and sustainability practices. This valuation underscores the intricate relationship between economic activities and natural systems, emphasising the imperative for the sustainable management of these resources. (Georgescu-Roegen, 1971; Bobulescu, 2012; Birch, 2016).

This line of research suggests that comprehending economic activities fully requires consideration of their social context, as economic systems operate as open subsystems within larger social and ecological systems. This underscores interconnectivity, where stability, integration, development, creation, linkage, and acculturation emerge as pivotal factors (Daly, 1993; Alikaeva et al., 2020, and Tedesco, 2022).

In light of the above discussion and in alignment with the Integration of Ecosystems Framework proposed by Tedesco (2022), the socio-economic conditions in which the studied ecosystems are situated are described below to provide additional context.

### 2.2.1 Oaxaca

Oaxaca is a state located in southern Mexico, with a population of approximately 4 million people, the majority of whom are indigenous (CONEVAL, 2022). Over 50% of Oaxaca's population speaks an indigenous language, making it one of the most culturally diverse states in the country (INEGI, 2020). The state is renowned for its rich cultural heritage, encompassing traditional music, dance, and cuisine.

Despite its cultural wealth, Oaxaca confronts several social challenges. Poverty looms large, with 66.5% of the population residing below the poverty line and 28.9% experiencing extreme poverty (CONEVAL, 2022). Indigenous communities and women bear a disproportionate burden, enduring higher poverty rates and facing limited opportunities for education and employment (INEGI, 2020). The state also faces a grave issue of violence against women, boasting one of the country's highest rates of femicides (INEGI, 2020).

Oaxaca's economy hinges predominantly on agriculture, with coffee, corn, and beans serving as the primary crops cultivated in the region (INEGI, 2020). Nevertheless, the state's economic vigour remains modest, with a per capita GDP of only $5,475 USD in 2020 (INEGI, 2020). Economic development has proven challenging, a reality underscored by the elevated poverty rates and prevailing inequality.

In addition to its social and economic tribulations, Oaxaca has also struggled with armed conflicts linked to its indigenous populace. Indigenous groups have long sought greater autonomy and recognition of their rights within the Mexican state (De Marinis, N., 2018). These conflicts have given rise to violence, displacement, and loss of life. The government has endeavoured to address these conflicts through legal and political avenues, yet the situation remains intricate and unresolved.

### 2.2.2 La Plata

La Plata serves as the capital city of Buenos Aires Province in Argentina, boasting a population of approximately 800,000 people (INDEC, 2022). The city maintains a relatively youthful demographic, with a median age of 33.5 years (INDEC, 2022).

La Plata also faces a host of societal challenges, including poverty, inequality, and violence. According to INDEC (2022), 29.3% of the population in Buenos Aires Province, which encompasses

La Plata, resides below the poverty threshold. Inequality emerges as a significant concern, with marginalized communities experiencing elevated poverty rates and encountering limited access to fundamental services like healthcare and education (Durante, 2019).

The city sustains a diversified economy, with a substantial portion of its GDP derived from the service sector, particularly education and healthcare services (INDEC, 2022). Concurrently, La Plata houses an informal agro-economy referred to as the "Cinturón flori-frutihortícola," denoting a region surrounding the city specializing in the production of flowers, fruits, and vegetables for local consumption and export (Fernández Acevedo, 2018).

The "Cinturón flori-frutihortícola" comprises an informal agro-economy that lacks government regulation and is often characterized by labour exploitation and substandard working conditions (Fernández Acevedo, 2018). A majority of the workforce in this sector consists of migrants, frequently from Bolivia and Paraguay, who confront discrimination, meagre wages, and limited access to essential services such as healthcare and education (Lemmi et al., 2020).

Notwithstanding the challenges faced by labourers in the "Cinturón flori-frutihortícola," this sector holds significant importance as a contributor to the local economy. It offers employment to thousands and generates substantial revenue through local sales and exports (Shoaie Baker & García, 2021). Nonetheless, the absence of regulation and labour protections within this sector has raised concerns regarding worker exploitation and the pressing need for improved working conditions (García, 2014).

### 2.2.3 Araucanía

Araucanía stands as a region located in southern Chile, with Temuco serving as its capital city. The region boasts a population exceeding one million individuals, with a notable proportion comprising indigenous Mapuche people (INE, 2017). The Mapuche people possess a unique culture and language and have confronted considerable historical injustices, giving rise to an enduring conflict that continues to impact the region (Ranjan et al., 2021).

The region has significant socio-economic challenges as Araucanía exhibits poverty rates surpassing the national average, accompanied by a persistent income inequality. Unemployment and informal labour arrangements prevail in the area, particularly among the Mapuche populace. Social issues, such as alcoholism, domestic violence, and youth delinquency further compound the region's challenges (ODEPA, 2018).

The Mapuche conflict in Araucanía holds a lengthy history, rooted in tensions stemming from the colonisation of Mapuche lands and the suppression of their culture and language. The conflict has taken various forms over the years, but recent manifestations involve violent attacks targeting forestry companies, law enforcement, and other governmental entities (Ranjan et al., 2021). The government's response has frequently drawn criticism for its heavy-handedness, exacerbating tensions and violence (Alberti et al., 2018).

Ongoing efforts have been made to address the social and economic challenges in Araucanía, alongside the Mapuche conflict. The government has implemented numerous programmes aimed at poverty reduction and economic development within the region. Additionally, initiatives have been launched to foster intercultural dialogue and understanding between Mapuche and non-Mapuche communities (ODEPA, 2018).

## 2. Data origin and pre-processing.

This research encompasses data collected in 2018 for the social innovation economic ecosystem in the city of Oaxaca, located in the Southeast region of Mexico; for the innovation-driven entrepreneurial economic ecosystem and the social-agricultural economic ecosystem in the City of La Plata in Buenos Aires, Argentina during 2021, and for the innovation-driven entrepreneurial economic ecosystem in the Araucanía region of southern Chile in 2022.

Data collection for each ecosystem commenced by initially identifying as many actors as possible through desk research. These identified actors were then invited to participate in a workshop aimed at strengthening innovation-driven entrepreneurial economic ecosystems. Additionally, they were requested to complete an online survey pertaining to their social interactions with other actors. The workshops and surveys were designed with four primary objectives, based on lean research principles (Hoffecker et al., 2015; Krystalli et al., 2021):

1. To gather quantitative and qualitative data concerning the relationships among actors.
2. To gather statistical data directly associated with collaboration outcomes between actors, irrespective of the outcomes' success, agreement nature, or other attributes corresponding to the social capital of each city.
3. To assist actors in defining their ecosystem's purpose, recognised as a fundamental element for complex systems, as elucidated by Meadows (2008).
4. To acquaint participants with the theoretical methodology for studying economic ecosystems and impart crucial lessons for enhancing and advancing their respective ecosystems.

At the heart of the research instrument, participants were asked to list up to 25 of their most relevant collaborations with other actors within the past three years and provide supplementary details regarding their nature and outcomes. All mentioned organisations were categorised in line with the TE-SER model, and the collaborations were transformed into network data, subsequently analysed and visualised using specialised software for complex network analysis—a process further elaborated on in Tedesco & Serrano (2019).

Following Tedesco's (2022) definition, we understand the grassroots innovation economic ecosystem as a community of actors that interact with each other and their environment within a delimited region, shaped by its social and natural dynamics, where resources are exchanged with the function of creating value to foster grassroot innovation.

Once the data was processed, a thorough examination of each organization's profile was conducted to identify actors pertaining to the grassroots innovation (GRI) economic ecosystem (Appendix 1), which play an indispensable role in generating innovative, bottom-up solutions to address social, economic, and environmental challenges within the realm of grassroots innovation. This examination involved scrutinising each organisation's purpose, description and the activities declared on their webpages and other available online profiles. In the context of this study, in accordance with Tedesco & Serrano (2019), we define a GRI actor as any organisation or initiative of an organisation that actively engages in the promotion, development, or implementation of social and sustainable innovations aimed at addressing issues pertaining to marginalised and/or vulnerable groups. As a consequence of this definition, the actors considered as part of this ecosystem in particular include not only grassroots organizations, but other supporting institutions as well. This is particularly applicable when referring to actors such as certain universities or governmental organizations which, despite not being grassroots organizations per-se, are in fact considered as actors due to their focus and declared initiatives in support of this ecosystem.

It is worth noting that the extensive participation of actors associated with the grassroots innovation economic ecosystem was not intentional. In each case, the initial focus was to study the larger innovation-driven entrepreneurial ecosystem. However, this phenomenon emerged as a characteristic feature due to the regional and cultural contexts of the studied economic ecosystems.

## 4. Findings

### 4.1 Brief analysis of the structural positioning of grassroot innovation actors.

For this publication, we revisit the data collected for the innovation-driven entrepreneurial economic ecosystems of Oaxaca, Araucanía, and La Plata, with a specific focus on the role and positioning of grassroots innovation actors identified as relevant based on Complex Network Analysis metrics (Tedesco, et al., 2022).

In the following graphs, referred to also as sociograms, a mathematical graphical representations of the structure of an economic ecosystem, all actors pertaining to the grassroots innovation economic ecosystem are marked in blue, while other actors from the overall innovation-driven entrepreneurial ecosystem are shown in grey.

By using a force-directed layout, these graphs provide visual representations of the resulting collaboration structure among actors within each economic ecosystem. Node size is dependent on the number of mentions by other participants and the strength or relevance of those mentions (weighted in-degree).

These visualizations, along with the interpretation of each node's centrality metrics, enable an analysis of the positioning of grassroots innovation actors within a larger economic ecosystem.

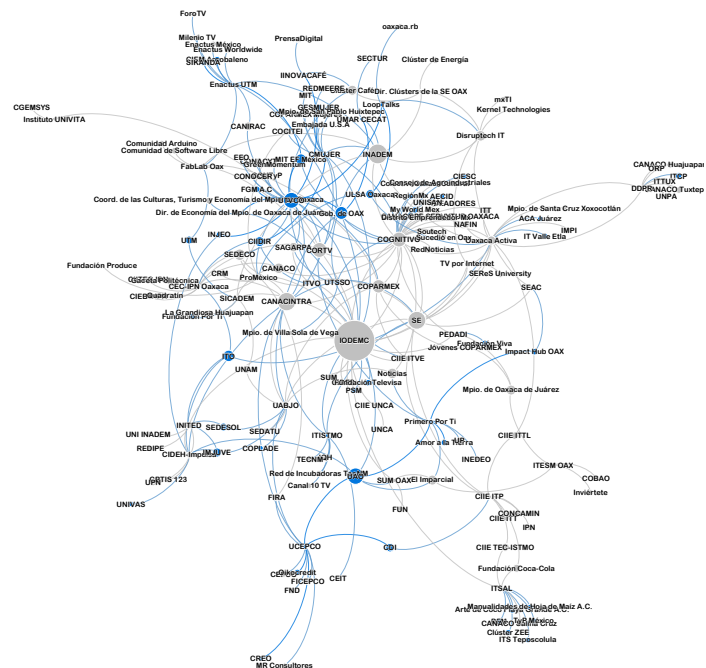

Figure 1. Structure of the Innovation-Driven Entrepreneurial Economic Ecosystem in Oaxaca, Mexico, including relationships with actors from the Grassroot Innovation ecosystem.

In the previous figure, all the organizations belonging to the grassroot innovation economic ecosystem, identified during the research process, are represented in blue. The figure also illustrates the degree of integration that these actors have within the social innovation economic ecosystem of that region.

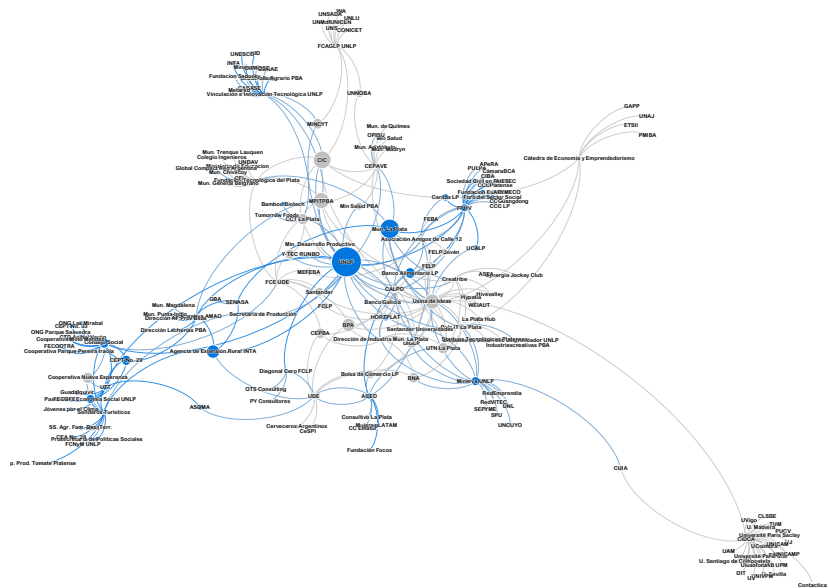

Figure 2. Structure of the Innovation-Driven Entrepreneurial Economic Ecosystem in La Plata, Argentina, Including Relationships with Actors from the Grassroot Innovation Ecosystem Related to the Regional Horticultural Economic Ecosystem.

In this other sociogram, it can be observed that due to the specific conditions of the economic ecosystem in the city of La Plata and the particularity associated with the horticultural belt of the city, which is mainly driven by the informal economy of migrant individuals, the actors of grassroots innovation not only find themselves at a higher level of integration but also exert greater influence, with the National University of La Plata playing a central role in the ecosystem.

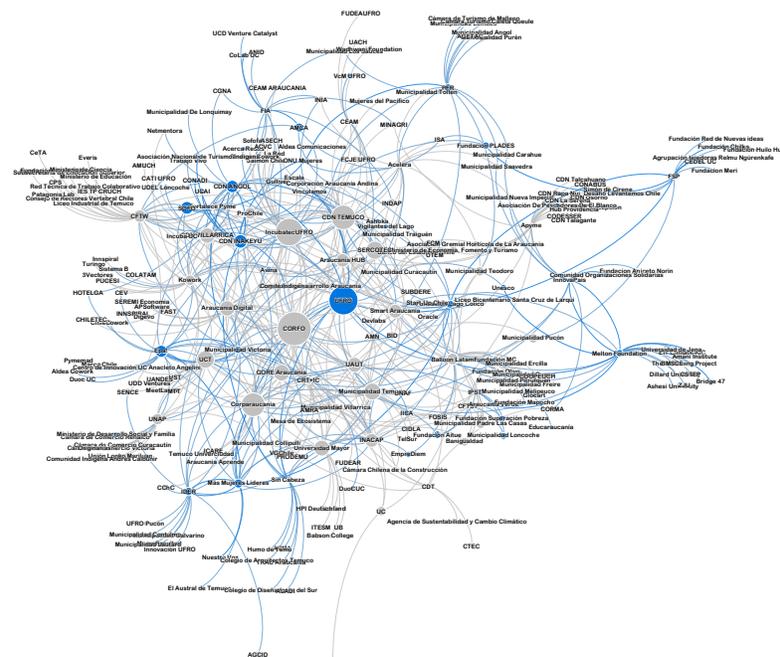

Figure 3. Structure of the Innovation-Driven Entrepreneurial Economic Ecosystem in Araucanía, Chile, including relationships with actors from the Grassroots Innovation ecosystem.

Finally, due to the complexity associated with the specific needs of the Araucanía region in the context of armed conflict, a greater participation of actors associated with the grassroot innovation ecosystem can be observed within the activities of the broader economic ecosystem of the region.

Similar to La Plata, there are prominent and central actors within the ecosystemic structure. The main actor in terms of weight and influence, as well as La Plata ecosystem, is a university called Universidad de La Frontera (UFRO).

The phenomenon of merging between ecosystems not only validates the *Ecosystems Integration Framework* (Tedesco & Serrano, 2019 and Tedesco, 2022), which theorizes about the overlapping between different economic sub-ecosystems, each with its own distinct functions, to form the overall economic ecosystem as a whole; it also demonstrates the necessary influence that actors from the grassroot innovation ecosystem have on regional economies with specific social vulnerabilities, whether it be an ecosystem shaped by a social context with a strong presence of indigenous peoples speaking over 14 active languages, in an ecosystem heavily influenced by an economy driven by vulnerable migrants, or in an ecosystem marked by socio-economic vulnerability and a long-standing armed conflict that has persisted for generations.

The following table provides further information on the behavior of GRI actors within their respective innovation-driven ecosystems according to their reported collaborations and some of their centrality metrics:

| Ecosystem | GRI participants as % of total sample | GRI nodes as % of total nodes in the ecosystem | % of collabs targeting GRIs | Average % of collabs from GRI participants targeting other GRIs | Difference in average weighted indegree of GRI nodes vs non-GRI nodes | Difference in average eccentricity of GRI participants vs non-GRI participants |
|---|---|---|---|---|---|---|
| Oaxaca | 33.3% | 29.4% | 27.0% | 20.0% | -10.5% | 0% |
| La Plata | 39.1% | 30.1% | 33.1% | 49.8% | 16.4% | 1.1% |
| Araucanía | 31.1% | 22.0% | 21.9% | 21.0% | 3.5% | 8.1% |

Table 1. Grassroot Innovation (GRI) actors nodal position analysis in the ecosystemic structure.

On average, Oaxaca's Grassroot Innovation actors are targeted slightly less often than expected, specially by other GRIs, and are perceived as slightly less relevant than non-GRIs as exhibited by their lower average weighted indegree. On the other hand, on average, La Plata Grassroot Innovation actors are targeted more often than expected for the percentage of GRI actors present in the ecosystem, specially by other GRIs, and are perceived as slightly more relevant than non-GRIs. And finally, despite the number of identified Grassroot Innovation actors in the Araucanía ecosystem, on average, GRIs are slightly more distant within the network as exhibited by their higher average eccentricty.

### 4.2 Roles and Values of Actors in the Grassroot Innovation Economic Ecosystem.

The following sociograms provide a visualization of the role and value (Tedesco & Serrano, 2019) that each grassroot innovation actor contributes to the ecosystemic structure. This adjacent value delivered by the actors of grassroot innovation represents the diversity that this ecosystem is capable of bringing to an even larger economic ecosystem, from which all users can greatly benefit. Colored edges represent collaborations directly mentioned by a participating grassroot innovation actor.

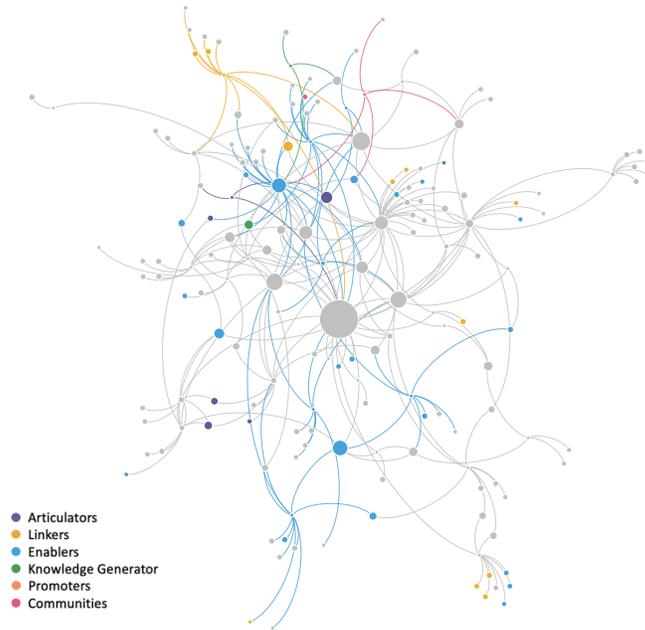

Figure 4. Actors of the Grassroot Innovation ecosystem by role and their relationships on Oaxaca, México

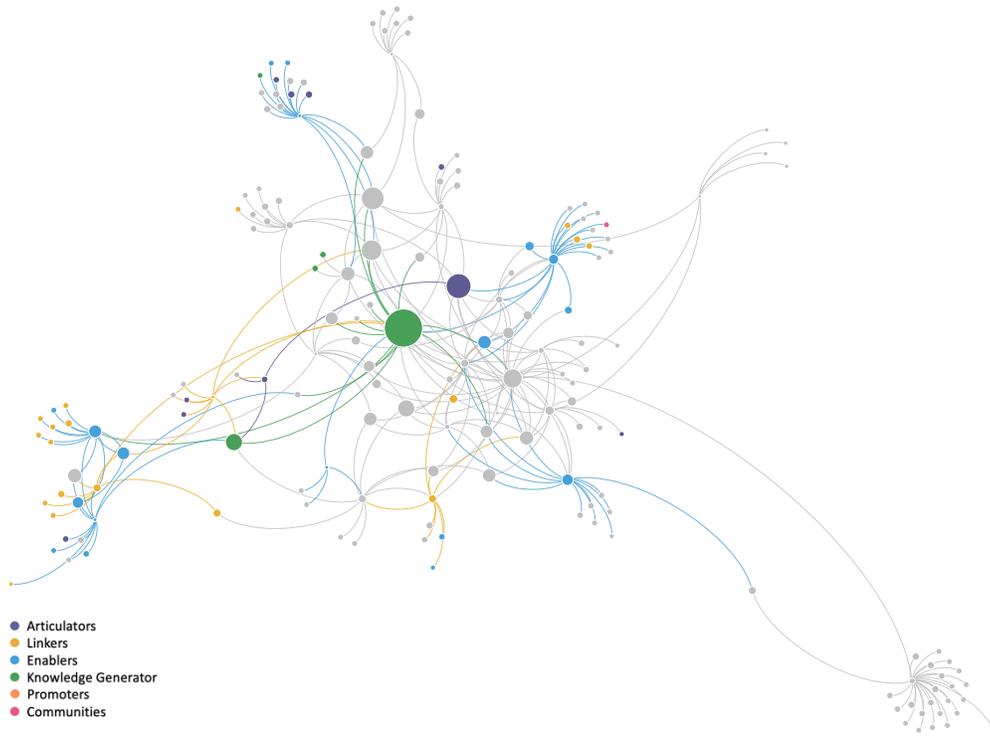

Figure 5. Actors of the Grassroot Innovation ecosystem by role and their relationships on La Plata, Argentina.

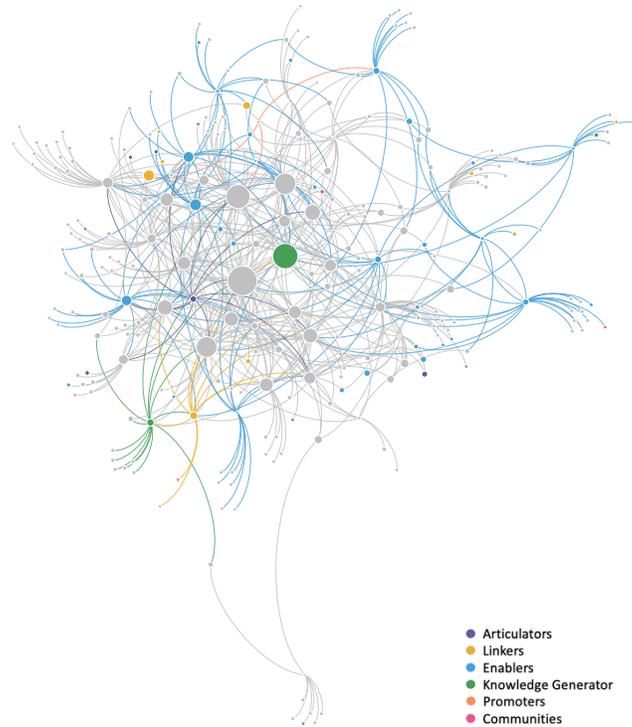

Figure 6. Actors of the Grassroot Innovation ecosystem by role and their relationships on Araucanía, Chile.

| Ecosystem | Articulators | Enablers | Linkers | Knowledge Generator | Promoters | Communities |
|---|---|---|---|---|---|---|
| Oaxaca | 12.0% | 54.0% | 24.0% | 6.0% | 0.0% | 4.0% |
| La Plata | 18.9% | 34.0% | 35.9% | 9.4% | 0.0% | 1.9% |
| Araucanía | 8.8% | 61.4% | 17.5% | 5.3% | 3.5% | 3.5% |

Table 2. Percentage of Grassroot Innovation (GRI) actors by role present in the ecosystem.

In general terms, the distribution of actors per role identified remains relatively consistent and comparable to the findings across other studies carried out using this approach (Tedesco, 2020a, b, c, d, e, f). However, certain deviations have emerged, warranting notable attention. Specifically, in the cases of Oaxaca and La Plata, no organization fulfilling the role of "Promoter" – a fundamental role in disseminating ecosystem stories and shaping its culture – was discernible based on the available data. This absence of actors actively championing and propagating the narratives of these ecosystems represents a distinct departure from the prevailing patterns observed thus far. While it is not uncommon for such actors to exhibit limited engagement in most studied ecosystems, the complete absence thereof is an unprecedented phenomenon.

Moreover, in the cases of Oaxaca and Araucanía, there is a discernible lack of participation from "Knowledge Generators" in comparison to other economic ecosystems. However, it is noteworthy that Araucanía holds a distinct advantage, akin to La Plata, as one of its key actors is a "Knowledge Generator," specifically the University of La Frontera (UFRO). This scarcity may impede the effective transfer of high-level innovation resources to these ecosystems, as these actors are responsible for generating high-value scientific or technological knowledge, including appropriate technology for vulnerable contexts, that can be utilized by innovators or entrepreneurs to promote sustainable development.

Conversely, in La Plata, a pronounced influence and active engagement of the government can be observed through the pivotal role of "Articulators," who are responsible for ensuring coherence in the economic ecosystem through public policy for the common good. On the other hand, since

"Articulators" are typically the type of role that tends to provide a wide range of resources, including financial support, training, incubation, or acceleration, as well as shared spaces, to facilitate the development of new innovations and ventures, the apparent scarcity of these actors raises concerns regarding the potential shortage of resources transferred from the larger innovation-driven economic ecosystem to the La Plata's GRI ecosystem and limiting access to these resources for grassroots innovators and entrepreneurs.

Furthermore, it is noteworthy that in Oaxaca and La Plata, there is a substantial percentage of "Linkers," a role often assumed by business organizations such as trade chambers, industrial councils, or formal associations of entrepreneurs. This underscores the strong presence of organizations that bring together diverse social groups to promote sector-specific public policies that advance the interests of these groups.

Finally, while "Communities," informal groups focused on disseminating knowledge, such as entrepreneurship clubs or innovation clubs, among others, tend to have a limited presence in terms of percentage within all economic ecosystems, they play an important role not only in dissemination but also in the dynamism of ecosystems and the representation of interests, as they can become formal groups over time. This could be particularly important in GRI ecosystems to address the needs of vulnerable groups.

## 5. Discussion and Future Research

The methodological approach developed in Tedesco (2022) and Tedesco et al., (2022) aims to understand the structural conditions of economic ecosystems based on collaboration dynamics among actors and how these structures reach equilibrium as competition is reduced and beneficial relationship mechanisms emerge (Wootton, JT & Emmerson, M., 2005; Le Roux J.J. et al., 2020). Social Network Mapping techniques and the TE-SER model implementation are highly useful in understanding how such structures also form in grassroots innovation ecosystems and the value these actors contribute to both their own ecosystems and the others they interact with.

Moreover, the data utilized in this study was not originally collected with the specific purpose of identifying the structures of grassroots innovation economic ecosystems. Nevertheless, it is noteworthy that, as an emergent phenomenon, this methodological approach facilitated the discovery of compelling evidence, enabling a comprehensive understanding of the influence and structural positioning of these actors. This further corroborates the efficacy and utility of the aforementioned approach within the realm of scientific inquiry.

The findings show that in all three innovation-driven entrepreneurial economic ecosystems, grassroots innovation actors play a significant role, as their participation in the study exceeded 30% of the total participants in all cases. Furthermore, in the cases of Oaxaca and La Plata, the total of grassroots innovation actors identified throughout the studied economic ecosystem represented approximately 30% of the total nodes as well. In comparison, it appears that in Araucanía, there is still work to be done to increase the participation of grassroots innovation actors in larger economic ecosystems that would allow them to leverage knowledge, infrastructure, and other relevant resources for the sustainability of their own economic ecosystem. The total of these actors represents just above 20% of the total in the studied ecosystem. This is particularly relevant considering the social and economic context of Araucanía and the concurrent issue known as the Mapuche conflict. Greater integration can result in social and economic benefits for both the more vulnerable population and the increased diversity in the innovation-driven economy as a whole.

Nevertheless, the structural network metrics show that grassroots innovation actors present in the three studied economic ecosystems are adequately integrated into a larger economic ecosystem, as they are only slightly more structurally distant within the network than the rest of the actors. This could simply be due to the unique role these actors play with their focus on social and non-profit innovation. That said, it's not a bad idea to conceive initiatives that place them centrally in the structure and provide them with a greater capacity to transfer resources to their own ecosystems.

Finally, La Plata seems to be the studied economic ecosystem where grassroots innovation actors are more dynamic and collaborative, while in the other two, increasing collaborative relationships between grassroots innovation actors and those of the larger economic ecosystem could result in a better position within the structure.

An in-depth inquiry would offer a more nuanced understanding by encompassing comprehensive visualizations and intricate structural data, shedding light on how actors within the grassroot innovation domain actively shape their own ecosystem structures. This enhanced comprehension would, in turn, enable the formulation of well-defined strategies aimed at fostering the development and growth of these structures to better serve the interests and needs of their users.

Advancing with similar studies in other regions of the planet with similar socio-economic contexts could help us better understand how integration dynamics between grassroots innovation ecosystems and larger ones occur, as well as allow for a deeper understanding of the mutual benefits that arise.

For future research, we recommend conducting more detailed analyses of strategies and policies that can foster greater integration and collaboration between grassroots innovation actors and other major economic ecosystems. This could include designing specific incentives, strengthening local capabilities, and promoting investment in innovative initiatives. Furthermore, it would be beneficial to delve into case studies in other regions with similar socioeconomic challenges to gain a better understanding of specific integration dynamics and the factors influencing the success of these collaborations.

Revisiting these economic ecosystems to understand how these policies and strategies have altered their complex structure would also be desirable to advance our understanding of how these dynamics works.

## 6. Limitations

Even though this analysis allows for a better understanding of how grassroots innovation actors interact within a larger ecosystem, since the original studies from which this data was derived were focused on studying innovation-driven entrepreneurial economic ecosystems and not the grassroots innovation ecosystem itself, it is currently incapable of fully describing the structural particularities of this sub-ecosystem.

In addition, despite a thorough examination of each organization's profile and following widely accepted definitions of what grassroots innovation entails, since very few organizations actually describe themselves as grassroots innovation actors, part of the classification process is ultimately subject to the authors' criteria – even though it is based on the characteristics declared by the actors – which might differ from that of other researchers.

# Appendix 1

## GRI Economic Ecosystem Actors Identified Per Region.

| Label | Full name | TE-SER Role | Study Participant |
|---|---|---|---|
| COPLADE | Coordinación de Planeación del Desarrollo | Articulator | No |
| Dir. de Economía del Mpio. de Oaxaca de Juárez | Dirección General de Economía del Municipio de Oaxaca de Juárez | Articulator | Yes |
| Gob. de OAX | Gobierno del Estado de Oaxaca | Articulator | No |
| IMJUVE | Instituto Mexicano de la Juventud | Articulator | No |
| INJEO | Instituto de la Juventud del Estado de Oaxaca | Articulator | No |
| SEDESOL | Secretaría de Desarrollo Social | Articulator | No |
| GESMUJER | Grupo de Estudios sobre la Mujer Rosario Castellanos, A. C. | Community | No |
| LoopTalks | LoopTalks | Community | Yes |
| AECID | Agencia Española de Cooperación Internacional para el Desarrollo | Enabler | No |
| Amor a la Tierra | En Amor a la Tierra | Enabler | No |
| CDI | Comisión Nacional para el Desarrollo de los Pueblos Indígenas | Enabler | No |
| CMUJER | Centro Universitario para el Liderazgo de la Mujer | Enabler | Yes |
| FGM A.C | Fondo Guadalupe Musalem A.C. | Enabler | No |
| Fundación Por Ti | Fundación Por Ti desde la Mixteca A. C. | Enabler | No |
| FundaciónTelevisa | Fundación Televisa | Enabler | No |
| Impact Hub OAX | Impact Hub Oaxaca | Enabler | No |
| IT Valle Etla | Instituto Tecnológico del Valle de Etla | Enabler | No |
| ITCP | Instituto Tecnológico de la Cuenca del Papaloapan | Enabler | No |
| ITISTMO | Instituto Tecnológico del Istmo | Enabler | Yes |
| ITO | Insituto Tecnológico de Oaxaca | Enabler | No |
| ITS Teposcolula | Instituto Tecnológico Superior de Teposcolula | Enabler | No |
| Manualidades de Hoja de Maíz A.C. | Asociación Civil de Manualidades de Hoja de Maíz de San Pedro Chanica | Enabler | No |
| My World Mex | My World México | Enabler | No |
| Oikocredit | Oikocredit | Enabler | No |
| Primero Por Ti | Primero por Ti | Enabler | Yes |
| PSM | Promotora Social México | Enabler | No |
| TyP México | Asociación Civil Pasión y Trabajo por un México mejor | Enabler | No |
| UAO | Universidad Anáhuac de Oaxaca | Enabler | No |
| UCEPCO | Unión de Crédito Estatal de Productores de Café de Oaxaca | Enabler | Yes |
| ULSA Oaxaca | Universidad Lasalle Oaxaca | Enabler | No |
| UMAR CECAT | Centro de Capacitación Turística de la Universidad del Mar | Enabler | Yes |
| UNIVAS | Universidad José Vasconcelos de Oaxaca | Enabler | No |
| UTM | Universidad Tecnológica de la Mixteca | Enabler | No |
| UTSSO | Universidad Tecnológica de la Sierra Sur de Oaxaca | Enabler | Yes |
| UTVCO | Universidad Tecnológica de los Valles Centrales de Oaxaca | Enabler | Yes |
| CIESC | Centro de Investigación y Estudios sobre Sociedad Civil A.C. | Knowledge generator | No |
| CIIDIR | Centro Interdisciplinario de Investigación para el Desarrollo Integral Regional Unidad Oaxaca | Knowledge generator | No |
| IINOVACAFÉ | Instituto de Investigación, Innovación y Adaptación del Café | Knowledge generator | Yes |
| ACA Juárez | Asociación de Cerveceros Artesanales de Oaxaca | Linker | No |
| Arte de Coco Playa Grande A.C. | Grupos Rural Asociación Civil Arte de Coco Playa Grande | Linker | No |
| BFM | Asociacion Civil de Artes de Bambú de San Miguel el Potrero | Linker | No |
| CIEM Arcobaleno | Centro Integrador Esperanza Mixteca Arcobaleno de México A.C | Linker | No |
| Clúster ZEE | Clúster de las Zonas Económicas Especiales | Linker | No |
| ColectivoOaxacaCultural | Colectivo Oaxaca Cultural | Linker | No |
| CREO | Centro de Regeneración Ecológica de Oaxaca | Linker | No |
| Enactus UTM | Enactus Universidad Tecnológica de la Mixteca | Linker | Yes |
| Fundación Viva | Fundación Viva | Linker | No |
| MIT EF México | MIT Entreprise Forum México | Linker | No |
| RegiónMx | Colectivo Región MX | Linker | No |
| SIKANDA | Solidaridad Internacional Kanda A.C | Linker | No |

Table 3. Organizations classified as grassroots innovation actor in Oaxaca.

| Label | Full Name | TE-SER Role | Study Participant |
|---|---|---|---|
| Desarrollo Agrario PBA | Ministerio de Desarrollo Agrario PBA | Articulator | No |
| Dirección AF Prov Bsas | Dirección Provincial de Agricultura Familiar y Desarrollo Rural | Articulator | No |
| Dirección Lecherías PBA | Dirección Lecherías Prov Bsas | Articulator | No |
| Industriascreativas PBA | Subsecretaría de Industrias creativas e innovación cultural | Articulator | No |
| Minhabitat | Ministerio de Desarrollo Territorial y Hábitat | Articulator | No |
| Minsocial | Ministerio de Desarrollo Social | Articulator | No |
| Mun. La Plata | Municipalidad de La Plata | Articulator | No |
| OPISU | Organismo Provincial de Integración Social y Urbana | Articulator | No |
| SENASA | Servicio Nacional de Sanidad y Calidad Agroalimentaria | Articulator | Yes |
| SS. Agr. Fam. Des. Terr. | Subsecretaría de Agricultura Familiar y Desarrollo Territorial | Articulator | No |
| AIESEC | AIESEC | Community | No |
| Banco Alimentario LP | Banco Alimentario La Plata | Enabler | No |
| BID | Banco Interamericano de Desarrollo | Enabler | No |
| Caritas LP | Caritas Arquidiocesana La Plata | Enabler | No |
| CEA No. 28 | Centro de Educación Agraria nº 28 | Enabler | No |
| CEPT No. 29 | Centro Educativo para la Producción Total N° 29 "Roberto Payró Escuela de Alternancia" | Enabler | Yes |
| CEPT No. 33 | Centro Educativo para la Producción Total N° 33 "El Deslinde" | Enabler | No |
| Consejo Social | Consejo Social de la Universidad Nacional de La Plata | Enabler | Yes |
| Diagonal Cero FCLP | Diagonal Cero FCLP | Enabler | Yes |
| FPHV | Fundación Pro Humanae Vitae | Enabler | Yes |
| Fundación Focos | Fundación Focos | Enabler | No |
| Minerva UNLP | Minerva UNLP | Enabler | Yes |
| Mujeres LATAM | Mujeres Latinoamericanas | Enabler | No |
| Paseo de la Economía Social UNLP | Paseo de la Economía Social UNLP | Enabler | No |
| Prosecretaría de Políticas Sociales | Prosecretaría de Políticas Sociales | Enabler | No |
| Senderos Turísticos | Programa de Extensión Universitaria Turismo, Patrimonio y Desarrollo en el periurbano platense, de la Universidad Nacional de La Pla | Enabler | Yes |
| UCALP | Universidad Católica de La Plata | Enabler | No |
| UNESCO | Organización de las Naciones Unidas para la Educación, la Ciencia y la Cultura | Enabler | No |
| Vinculación e Innovación Tecnológica UNLP | Secretaría de Vinculación e Innovación Tecnológica UNLP | Enabler | Yes |
| Agencia de Extensión Rural INTA | Agencia de Extensión Rural INTA | Knowledge Generator | No |
| Bamboo Biotech | Bamboo Biotech | Knowledge Generator | No |
| INTA | Instituto Nacional de Tecnología Agropecuaria | Knowledge Generator | No |
| Tomorrow Foods | Tomorrow Foods | Knowledge Generator | No |
| UNLP | Universidad Nacional de la Plata | Knowledge Generator | Yes |
| ACED | Asociación Civil Empresarias de las Diagonales | Linker | Yes |
| ASOMA | Asociación de Medieros | Linker | No |
| Coop. Prod. Tomate Platense | Cooperativa de Productores de Tomate Platense | Linker | No |
| Cooperativa AMAO | Cooperativa AMAO amanecer organizado | Linker | Yes |
| Cooperativa Moto Mendez | Cooperativa Moto Mendez | Linker | No |
| Cooperativa Parque Pereira Iraola | Cooperativa Parque Pereira Iraola | Linker | No |
| CTD Aníbal Verón | CTD Aníbal Verón | Linker | No |
| FECOFE | Federación de Cooperativas Federadas | Linker | No |
| FECOOTRA | Federación de Cooperativas de Trabajo de la República Argentina | Linker | No |
| Foro del Sector Social | Foro del Sector Social | Linker | No |
| Fundacion Eurosur | Fundación Eurosur | Linker | No |
| Global Compact Red Argentina | Global Compact Red Argentina | Linker | No |
| Guadalquivir | Asociación de Productores y Familiares El Guadalquivir | Linker | No |
| HORTPLAT | Asociacion de Productores Hortícola de La Plata | Linker | No |
| Jóvenes por el Clima | Jóvenes por el Clima | Linker | No |
| ONG Las Mirabal | ONG Las Mirabal | Linker | No |
| ONG Parque Saavedra | ONG Parque Saavedra | Linker | No |
| Sociedad Civil en Red | Sociedad Civil en Red | Linker | No |
| UTT | Unión de Trabajadores de la Tierra | Linker | Yes |

Table 4. Organizations classified as grassroots innovation actor in La Plata

| Label | Full name | TE-SER Role | Participant |
|---|---|---|---|
| CONADI | Corporación Nacional de Desarrollo Indígena | Articulator | No |
| Ministerio de Desarrollo Social y Familia | Ministerio de Desarrollo Social y Familia | Articulator | No |
| Municipalidad Loncoche | Municipalidad de Loncoche | Articulator | No |
| Municipalidad Victoria | Municipalidad de Victoria | Articulator | Yes |
| UDEL Loncoche | Unidad de Desarrollo Económico Local de Loncoche | Articulator | No |
| Bridge 47 | Bridge 47 | Community | No |
| Vigilantes del Lago | Vigilantes del Lago | Community | No |
| Acerca Redes | Acerca Redes | Enabler | No |
| Amani Institute | Amani Institute | Enabler | No |
| Araucania Aprende | Fundacion Araucania Aprende | Enabler | No |
| Araucanía Verde | Fundación Araucanía Verde | Enabler | No |
| Ashoka | Ashoka | Enabler | No |
| Avina | Fundación Avina | Enabler | No |
| CDN ANGOL | Centro de Negocios Sercotec Angol | Enabler | Yes |
| CDN INAKEYU | Centro de Negocios Inakeyu | Enabler | Yes |
| CDN Rapa Nui | Centro de Negocios Sercotec Rapa Nui | Enabler | No |
| CoLab UC | Laboratorio de Innovación Social de la Pontificia Universidad Católica de Chile | Enabler | No |
| Comité Indígena | Comité de Desarrollo y Fomento Indígena | Enabler | No |
| DuoCUC | DEPARTAMENTO UNIVERSITARIO OBRERO CAMPESINO de la Universidad Católica de Chile. | Enabler | No |
| Educaraucanía | Fundación Educa Araucanía | Enabler | No |
| EmpreDiem | EmpreDiem | Enabler | No |
| Epa! | Epa! Cowork Villarrica | Enabler | Yes |
| FIA | Fundación para la Innovación Agraria | Enabler | Yes |
| FSP | Fundacion Sustenta Pucón | Enabler | Yes |
| Fundación Aitue | Fundacion De Desarrollo Social Y Cultural Aitue | Enabler | No |
| Fundacion Lago Colico | Fundacion Lago Colico | Enabler | Yes |
| Fundación Mapocho | Fundación Mapocho | Enabler | No |
| Fundación MC | Fundación MC | Enabler | No |
| Fundación PLADES | Fundación PLADES Frutillar | Enabler | No |
| Fundación Superación Pobreza | Fundación Superación de la Pobreza | Enabler | No |
| Glocart | Glocart | Enabler | No |
| InnovaPaís | Fundación InnovaPaís | Enabler | Yes |
| Melton Foundation | Melton Foundation Temuco | Enabler | Yes |
| ONU Mujeres | ONU Mujeres | Enabler | No |
| Originarias | Originarias - ONU Mujeres | Enabler | No |
| PER | Programa Estratégico Regional de Turismo, Cultura y Naturaleza de Nahuelbuta y Araucanía Costera | Enabler | Yes |
| Sin Cabeza | Agencia Cooperativa Sin Cabeza | Enabler | Yes |
| Sistema B | Sistema B | Enabler | No |
| Unesco | Unesco | Enabler | No |
| VcM UFRO | Vinculación con el Medio Universidad de la Frontera | Enabler | No |
| Vinculamos | Vinculamos | Enabler | Yes |
| VIVA Idea | VIVA Idea | Enabler | No |
| CEDEL UC | Centro UC de Desarrollo Local | Knowledge Generator | No |
| IDER | Instituto de Desarrollo Local y Regional | Knowledge Generator | Yes |
| UFRO | Universidad de la Frontera | Knowledge Generator | No |
| AMCA | Asociación de Municipios Costa Araucanía | Linker | No |
| Asociación De Pescadores De El Blanco | Asociación Gremial De Pescadores Artesanales De Las Caletas El Blanco | Linker | No |
| Asociación Nacional de Turismo Indígena | Asociación Nacional de Turismo Indígena | Linker | No |
| Comunidad Organizaciones Solidarias | Comunidad Organizaciones Solidarias | Linker | No |
| Fundación Chilka | Fundación Chilka | Linker | No |
| IICA | Instituto Interamericano de Cooperación para la Agricultura | Linker | No |
| Más Mujeres Lideres | Más Mujeres Lideres | Linker | Yes |
| Mesa de Ecosistema | Mesa de Ecosistema de la Araucanía | Linker | No |
| SOFO | Sociedad de Fomento Agrícola de Temuco | Linker | No |
| UCAI | Unidad de Coordinación de Asuntos Indígenas | Linker | No |
| Aldea Comunicaciones | Aldea Comunicaciones | Promoter | Yes |
| Nuestra Voz | Nuestra Voz | Promoter | No |

Table 5. Organizations classified as grassroots innovation actors in Araucanía.